\documentclass[conference,a4paper]{APSIPA2021}
\IEEEoverridecommandlockouts
\usepackage{cite}
\usepackage{amsmath,amssymb,amsfonts}
\usepackage{algorithmic}
\usepackage{graphicx}
\usepackage{textcomp}
\usepackage{xcolor}
\usepackage{threeparttable}
\usepackage{multirow}
\usepackage{hyperref}

\def\BibTeX{{\rm B\kern-.05em{\sc i\kern-.025em b}\kern-.08em
    T\kern-.1667em\lower.7ex\hbox{E}\kern-.125emX}}
\begin{document}

\title{HouseX: A Fine-Grained House Music Dataset and Its Potential in the Music Industry\\
}
\author{%
\authorblockN{%
Xinyu Li\authorrefmark{1}
}
\authorblockA{%
\authorrefmark{1}
New York University Shanghai, Shanghai, China \\
E-mail: xl3133@nyu.edu  Tel/Fax: +86-19512484597}
}

\maketitle

\begin{abstract}
Machine sound classification has been one of the fundamental tasks of music technology. A major branch of sound classification is the classification of music genres. However, though covering most genres of music, existing music genre datasets often do not contain fine-grained labels that indicate the detailed sub-genres of music. In consideration of the consistency of genres of songs in a mixtape or in a DJ (live) set, we have collected and annotated a dataset of house music that provide 4 sub-genre labels, namely future house, bass house, progressive house and melodic house. Experiments show that our annotations well exhibit the characteristics of different categories. Also, we have built baseline models that classify the sub-genre based on the mel-spectrograms of a track, achieving strongly competitive results. Besides, we have put forward a few application scenarios of our dataset and baseline model, with a simulated sci-fi tunnel as a short demo built and rendered in a 3D modeling software, with the colors of the lights automated by the output of our model.
\end{abstract}


\section{Introduction}
One of the fundamental tasks of music information retrieval is the classification of music. For the task of audio classification, one of the most popular methods is to classify the spectrograms or mel-spectrograms of sounds using computer vision models like convolutional neural networks (CNN) \cite{cnn} or vision transformers (ViT) \cite{vit}. Jaiswal et al. \cite{sound_cls_cnn} has proposed the CNN-based approach that significantly improved the performance of sound classification. Researchers have delicately annotated datasets like the GTZAN \cite{gtzan} and the FMA \cite{fma_dataset}, etc. In view of the popularity of electronic dance music, we have collected a dataset that specifically features house music, which is the primary focus of production by many world-known DJ/producers. Our dataset consists of 160 labelled tracks covering 4 popular sub-genres, which are future house, bass house, progressive house and melodic house. We have also manually annotated the start and the end of the drop of each track. Besides, we have fine-tuned several CNN architectures pretrained on ImageNet \cite{imagenet} and trained a vanilla vision transformer, as baseline models for classification on this dataset. These models can achieve strongly competitively classification results on our test set. Finally, we have proposed two real-world scenarios that could benefit from our dataset and baseline model. The dataset along with the baseline models can be accessed via this \href{https://github.com/Gariscat/HouseX}{\color{blue}{link}}. For conciseness, in tables and figures, we denote future house by "FH", bass house by "BH", progressive house by "PH" and melodic house by "MH". In all, our main contributions are:

\begin{itemize}
    \item To the best of our knowledge, we have proposed the first house music dataset with sub-genre labels and drop annotations.
    \item We have provided a bunch of fine-tuned models on the task of sub-genre classification on this dataset.
    \item We have proposed a few real-world scenarios where our dataset and models could be applied. With a demo project presented, we have brought forward a new paradigm for automated music visualization.
\end{itemize}

\section{Related Works}

\subsection{Music Genre Datasets}
In the past decades, previous researchers annotated a few music genre datasets. Among them, one of the most widely used is called GTZAN \cite{gtzan}. Its audio data includes 10 genres, with 100 audio files each genre, all having a duration of 30 seconds. Besides, they have also provided the converted mel-spectrograms, together with the mean and variance of multiple acoustic features, including the spectral centroid and the root mean square (RMS). Defferrard et al. have built the FMA dataset \cite{fma_dataset} which includes 8 genres and is also annotated in the style of GTZAN. Other works like the MedleyDB dataset \cite{medleyDB} contains classical, rock, world/folk, fusion, jazz, pop, musical Theatre and rap songs. Besides genre labels, it also provides processed stems for its 194 tracks.
\subsection{Limitations}
However, most music genre datasets, including the ones mentioned in the last paragraph, are annotated with general genres like pop, country, rock, etc. One shortcoming is that models trained on these datasets could hardly distinguish the differences among songs within a general genre. In one mixtape or one live set, the songs usually belong to one or few general genres. For example, all the tracks selected in our dataset would be classified as "Electronic" in datasets like the FMA dataset \cite{fma_dataset}. Therefore, using models trained on these datasets, the retrievable information would be extremely confined. In view of such limitation, our work is devoted to building datasets that feature electronic music belonging to the same general genre but differ in their sub-genres.

\section{Approaches}
In this section, we discuss how we process the data, and the classification task in a rigorous, mathematical format. Also, we would present our deep learning approach, including the pipeline of setting up baseline models.

\subsection{Data Annotation}\label{AA}
After our evaluation on its wide usage on music festivals, we have chosen house music as the general genre of our dataset, naming it "HouseX". The dataset includes 4 sub-genres of house music, with 40 tracks each. These 160 tracks are all produced by the most popular DJ/producers around the world, including \cite{garrix,tobu}.

\addtolength{\topmargin}{0.001cm}
\begin{table}[!tb]
\renewcommand\arraystretch{1.25}
\setlength{\tabcolsep}{4mm}
\centering
\caption{Basic Statistics of HouseX}
\label{tab:stats}
\scalebox{0.9}{
\begin{tabular}{ll}
\hline\hline
Number of Tracks (train) & 128 \\
Number of Tracks (val) & 20 \\
Number of Tracks (test) & 12 \\
Number of Mel-spectrograms (train) & 13932 \\
Number of Mel-spectrograms (val) & 2184 \\
Number of Mel-spectrograms (test) & 1364 \\
Number of Mel-spectrograms (train, drops only) & 2183 \\
Number of Mel-spectrograms (val, drops only) & 339 \\
Number of Mel-spectrograms (test, drops only) & 204 \\
\hline
BPM Range        & 122-130 \\
Total Length (HH:MM:SS)     & 09:07:54 \\
\hline\hline
\end{tabular}}
\end{table}

\begin{figure}[!tb]
    \centering
    \includegraphics[width=0.47\textwidth]{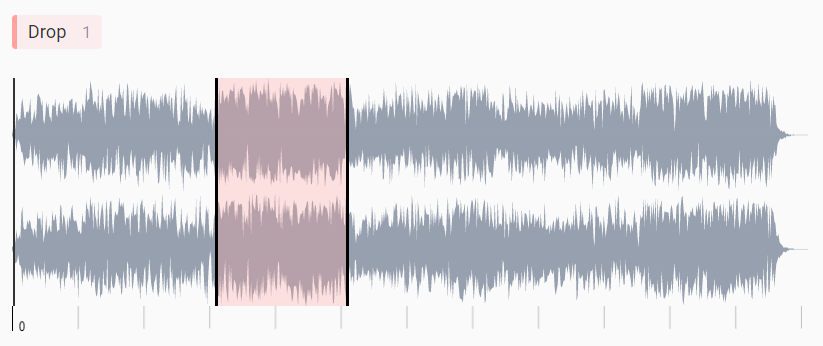}
    \caption{An Example of Annotating the Drop}
    \label{fig:drop}
\end{figure}

Though having similar or even identical tempos, these four sub-genres of house music possess substantially different characteristics. Future house tracks include plenty of electronic sounds, and their groove can be very bouncy (note that a major sub-branch of future house is future bounce). Bass house music is also filled with synthesized electronic elements. However, the sounds in bass house are generally dirtier and more distorted. These tracks could be so aggressive that one can hardly discern a clear melody line. Though sometimes bass house could sound similar to future house, future house tracks usually have a bright lead sound while bass house would possess mush more richness and saturation on the low end of the frequency spectrum. Progressive house is known to be uplifting and emotional. Beside the widely used supersaw lead sound, many progressive house producers would use real instruments like the piano, the guitar and strings to make the sounds more natural and organic. Melodic house tracks are characterized by the use of pluck sounds and pitch shifting techniques. These elements produce a happy and rhythmical atmosphere.

We have counted a few statistics of our dataset. The tempo of these $160$ tracks range from $122$ BPM to $130$ BPM, and most of them equal $128$. The total duration of them is around $9$ hours and $8$ minutes. Most house tracks have a time signature of $4$/$4$. Therefore, suppose a track has a tempo of $128$ bpm, then one bar in the track has a duration of $1.875$ seconds. We set the length of the sound clip for each mel-spectrogram as $1.875$ seconds, which are is approximately $1$ bars for every track. In total, we derive around $17480$ mel-spectrograms for classification. Some detailed statistics are presented in Tab. \ref{tab:stats}.

Additionally, we have annotated the timestamps that indicate the start and the end of the drop of each track, using Label Studio \cite{LabelStudio}. Since the drop is the climax of a track, with respect to energy and emotion, it is the section where the characteristics of its sub-genre are best exhibited. For example, the metallic lead sound in the drop and heavily distorted bass would characterize a bass house track, while the supersaw lead layered with strings and piano would highlight an emotional melody line of a progressive house track. Therefore, we specifically annotated the start and the end of the drop of each track in our dataset, as shown in Fig. \ref{fig:drop}. These drops would form a subset of the original dataset. In order to make this subset balanced, we only choose one drop with a length of $16$ bars, from each track. Normally, a track includes two drops. We choose the first one unless it is not long enough. Sometimes, especially for emotional progressive house tracks, the first drop only contains $8$-bars while the length of the second one is doubled. In these cases, we annotate the start and the end of the second drop. Though fairly smaller than the whole set, this subset that only contains drops still yields a considerable amount of mel-spectrograms, as listed in Tab. \ref{tab:stats}.
Another thing to note is that the split of the whole set should be done before the conversion from audio to mel-spectrograms. That is to say, we should split the set of tracks, rather than the set of pictures, into $3$ parts, for training, validation and testing respectively. The reason is that electronic music usually contains many loops, and if we split the set of pictures, for a mel-spectrogram in the test set, similar mel-spectrograms would almost surely appear in the training set, which makes the validation and testing process meaningless.

\begin{figure*}[!htp]
    \centering
    \includegraphics[width=.24\textwidth]{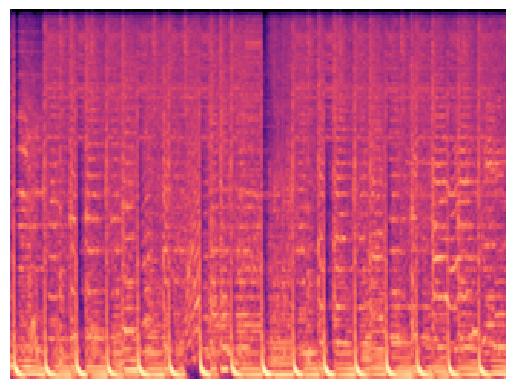}
    \includegraphics[width=.24\textwidth]{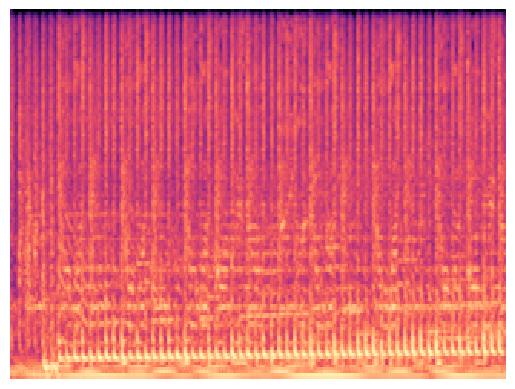}
    \includegraphics[width=.24\textwidth]{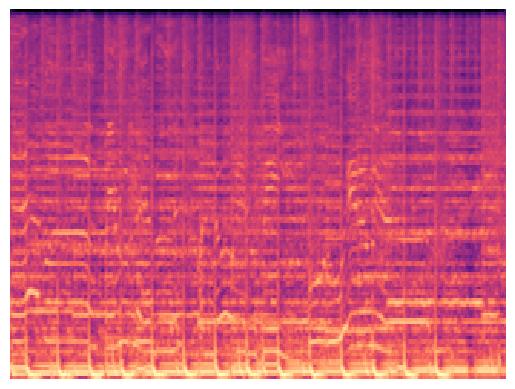}
    \includegraphics[width=.24\textwidth]{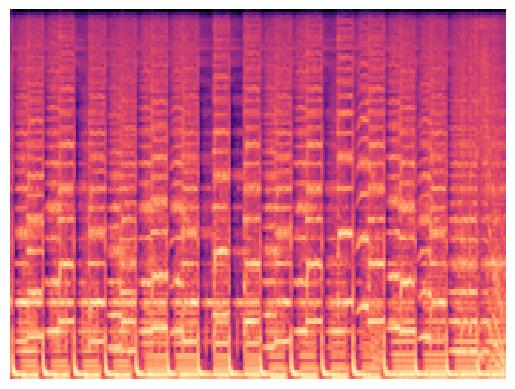}
\caption{Examples of Mel-spectrograms in the Dataset ($FH$, $BH$, $PH$, $MH$ from Left to Right)}
\label{fig:mel-ex}
\end{figure*}

\begin{figure}[!htp]
    \centering
    \includegraphics[width=9cm]{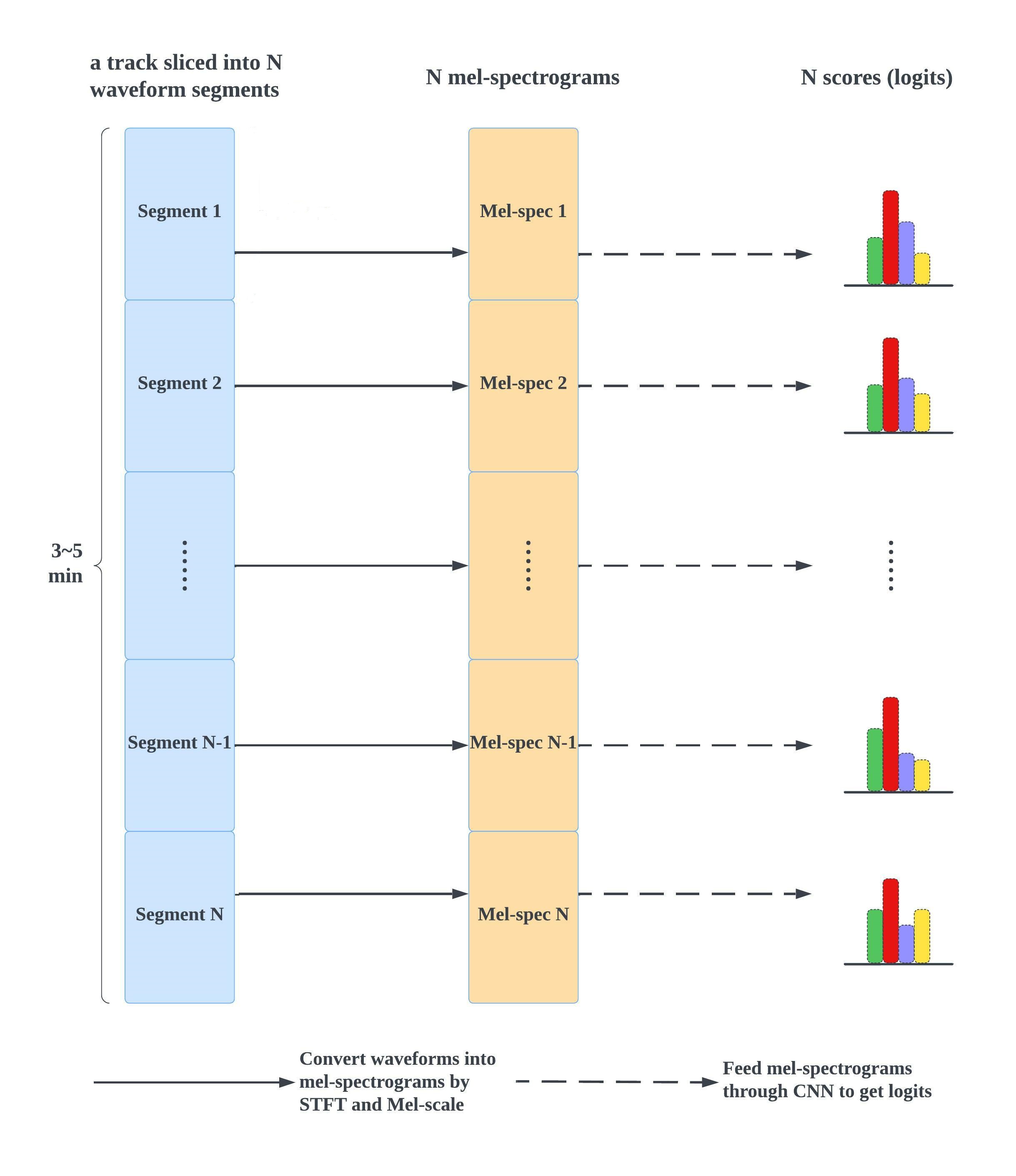}
    \caption{The Pipeline of the Classification Task}
    \label{fig:pipeline}
\end{figure}

\begin{figure*}[!htbp]
    \centering
    \includegraphics[width=.23\textwidth]{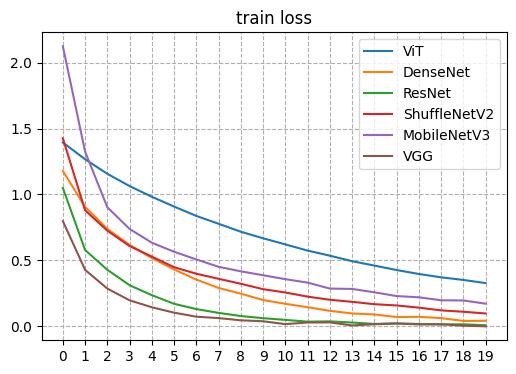}
    \includegraphics[width=.23\textwidth]{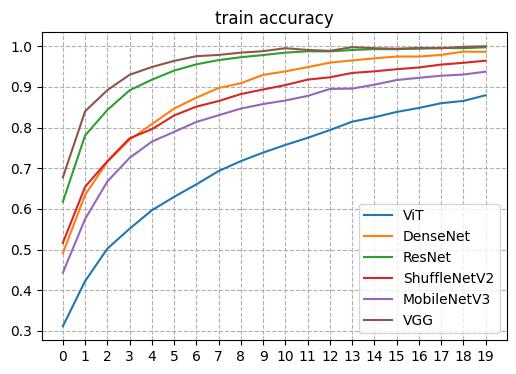}
    \includegraphics[width=.23\textwidth]{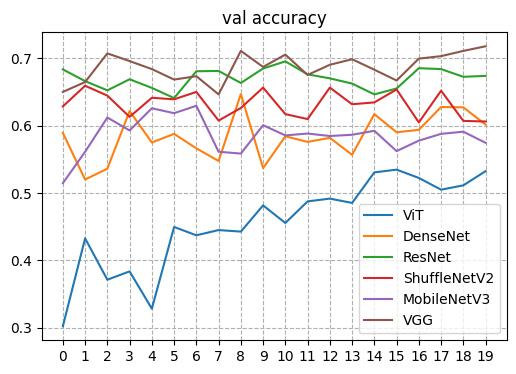}
    \includegraphics[width=.23\textwidth]{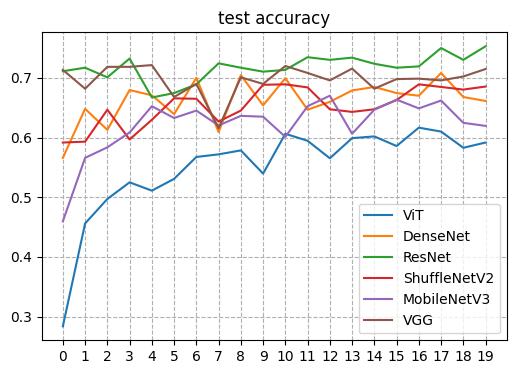}
\caption{The Learning Curves on the Whole Set ($x$-axis indicates the number of epochs)}
\label{fig:logs_ori}
\end{figure*}
\begin{figure*}[!htbp]
    \centering
    \includegraphics[width=.23\textwidth]{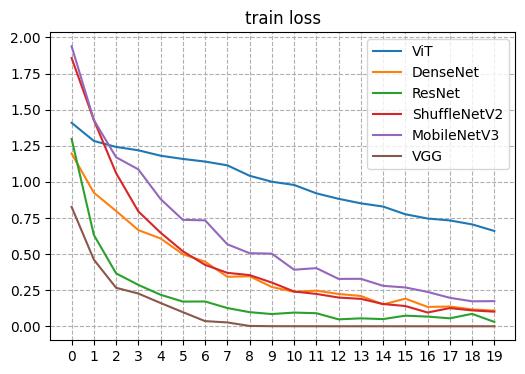}
    \includegraphics[width=.23\textwidth]{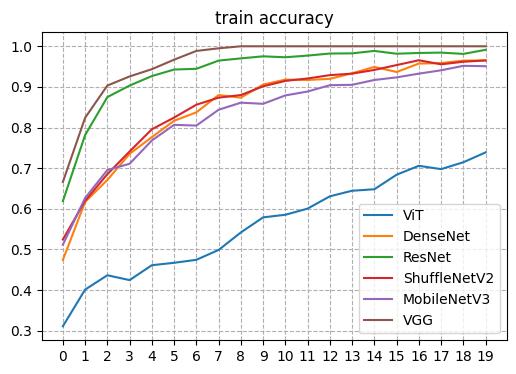}
    \includegraphics[width=.23\textwidth]{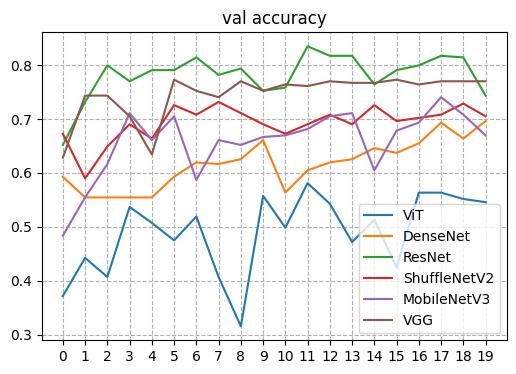}
    \includegraphics[width=.23\textwidth]{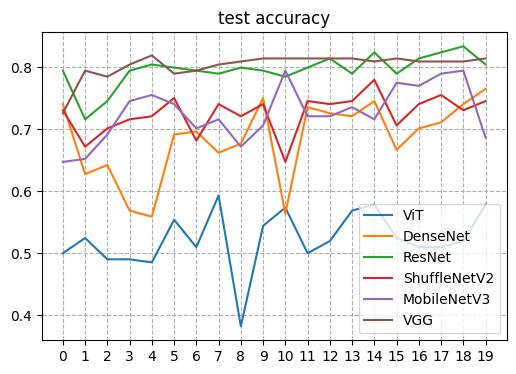}
\caption{The Learning Curves on the Drops-only Subset ($x$-axis indicates the number of epochs)}
\label{fig:logs_sub}
\end{figure*}

\subsection{Task Definition}
Rigorously, the classification task is defined as follows. A track of house music is given as a wave file and the goal is to assign a proper sub-genre to slices of the track, from the $4$ sub-genres mentioned in the data collection part. Since our wave files are stereo, for each track, we get an array of shape $(2, T)$ where $T$ is the total number of samples, and the sample rate of the waveform $sr=22050$ (Hz). To get reasonable amount of mel-spectrograms, we set the number of bars per slice as $N_{bar\_per\_slice}=1$. Then, since all tracks in our dataset have a tempo of $128$ BPM (or close to $128$ BPM) and a time signature of $4/4$, we have $$L=\lfloor{\frac{4\times 60\times N_{bar\_per\_slice}\times sr}{128}}\rfloor\approx 41343$$, as the number of samples for each slice. Then, without overlapping, we cut the track into $N=\lfloor\frac{T}{L}\rfloor$ slices, each as an array with shape $(2, L)$. Using functions implemented in Librosa \cite{librosa}, we convert these $N$ Numpy arrays into $N$ mel-spectrograms, each with a shape of $$(n\_mels, \lceil\frac{L}{hop\_length}\rceil)$$, where $n\_mels=128$ and $hop\_length=512$ as default. After saving these mel-spectrograms rendered by the $specshow$ function of Librosa, we can load them as JPEG files, transform them into PyTorch tensors \cite{pytorch} and proceed with the standard image classification workflow. A few saved mel-spectrograms are shown in Fig. \ref{fig:mel-ex} as examples.

\newcommand{\comment}[1]{}
\comment{
That is to say, we resize the images to $(3, 224, 224)$ tensors and feed batches of them (each batch has a shape of $(bs, 3, 224, 224)$ and the batch size $bs=4$ as default) through the pretrained network to get batches of logits (each batch of logits is a tensor with shape $(bs, 4)$). For each batch, the labels are represented as a tensor with shape $(bs, )$, where the $i$-th element in the tensor indicates the sub-genre of the $i$-th mel-spectrogram, or precisely, the waveform slice corresponding to that mel-spectrogram. For simplicity, we assume that slices cut from the same track have the same sub-genre as the original track.
}

\begin{table*}[!htbp]
    \begin{center}
    \begin{threeparttable}
    \caption{Experiment Results on the Validation Set and the Test Set}
    \label{tab:results}
    \begin{tabular}{|c|c|c|c|c|c|c|c|c|c|c|c|c|}
    
    \hline
    
    \multirow{3}{6mm}{\centering Model} & \multicolumn{6}{c|}{Val} & \multicolumn{6}{c|}{Test}\\
    
    \cline{2-13}
    
     & \multirow{2}{8mm}{\centering Acc.\tnote{a}\quad (\%)} & \multicolumn{5}{c|}{\centering F1-score} & \multirow{2}{8mm}{\centering Acc.\tnote{a}\quad (\%)} & \multicolumn{5}{c|}{\centering F1-score} \\
     
    \cline{3-7}\cline{9-13}
    
     & & FH & BH & PH & MH & Weighted Avg & & FH & BH & PH & MH & Weighted Avg \\
     
    \hline
    
    MobileNetV3 &57.46&0.4&0.67&0.55&0.64&0.57&61.95&0.37&0.71&0.62&0.68&0.6\\
    \hline
    \textbf{ResNet18}\tnote{b} &67.4&0.57&0.76&0.67&0.69&0.68&75.29&0.58&0.79&0.83&0.78&0.75\\
    \hline
    VGG16 &71.79&0.64&0.74&0.74&0.75&0.72&71.48&0.47&0.79&0.79&0.75&0.71\\
    \hline
    DenseNet121 &60.16&0.51&0.71&0.53&0.64&0.6&66.13&0.51&0.72&0.67&0.69&0.65\\
    \hline
    ShuffleNetV2 &60.62&0.52&0.63&0.63&0.64&0.61&68.55&0.49&0.79&0.75&0.69&0.69\\
    \hline
    (Vanilla) ViT &53.25&0.4&0.56&0.5&0.63&0.53&59.02&0.45&0.71&0.53&0.63&0.58\\
    
    \hline
    \end{tabular}
    
    \begin{tabular}{|c|c|c|c|c|c|c|c|c|c|c|c|c|}
    
    \hline
    
    \multirow{3}{6mm}{\centering Model} & \multicolumn{6}{c|}{Val (Drops Only)} & \multicolumn{6}{c|}{Test (Drops Only)}\\
    
    \cline{2-13}
    
     & \multirow{2}{8mm}{\centering Acc.\tnote{a}\quad (\%)} & \multicolumn{5}{c|}{\centering F1-score} & \multirow{2}{8mm}{\centering Acc.\tnote{a}\quad (\%)} & \multicolumn{5}{c|}{\centering F1-score} \\
     
    \cline{3-7}\cline{9-13}
    
     & & FH & BH & PH & MH & Weighted Avg & & FH & BH & PH & MH & Weighted Avg \\
     
    \hline
    
    MobileNetV3 &66.96&0.61&0.77&0.64&0.63&0.66&68.63&0.59&0.73&0.78&0.62&0.68\\
    \hline
    \textbf{ResNet18}\tnote{b} &74.34&0.68&0.76&0.85&0.68&0.74&80.39&0.7&0.82&0.93&0.74&0.8\\
    \hline
    VGG16 &76.99&0.74&0.8&0.77&0.77&0.77&81.37&0.66&0.83&0.92&0.83&0.81\\
    \hline
    DenseNet121 &69.62&0.66&0.75&0.72&0.65&0.7&76.47&0.72&0.8&0.8&0.74&0.76\\
    \hline
    ShuffleNetV2 &70.5&0.64&0.76&0.78&0.65&0.71&74.51&0.62&0.77&0.87&0.71&0.75\\
    \hline
    (Vanilla) ViT &54.57&0.35&0.79&0.42&0.54&0.52&57.84&0.58&0.61&0.72&0.32&0.56\\
    
    \hline
    \end{tabular}
    
    \begin{tablenotes}
    \item[a] The accuracy scores are calculated on the entire validation/test set, over all categories.
    \item[b] In the demo of the first proposed application, we choose ResNet18 \cite{resnet} due to its high accuracy and relatively small number of parameters, which leads to comparatively fast inference speed.
    \end{tablenotes}
    \end{threeparttable}
    \end{center}
\end{table*}

\subsection{Pipeline}
As shown in the Fig. \ref{fig:pipeline}, the pipeline consists of two major components. The first step is the conversion from sliced sound waveforms to mel-spectrograms, and the second procedure is to feed mel-spectrograms through a neural network to get the output scores for these mel-spectrograms. Each output score is a $(4, )$ vector.

In the training phase, we iterate through the training set to minimize the negative log likelihood loss between the output logits and the label tensor. Denoting $\mathcal{D}$ as the dataset and $\mathcal{C}$ as the set of labels, mathematically, we need to minimize the objective
$$\mathcal{J}(\theta) = -\sum_{(x, y)\in \mathcal{D}}\sum_{c\in\mathcal{C}}{\log(p_{\theta}(c|x))\mathbb{I}_{y==c}}$$
, where $\theta$ represents the parameters of the model, each $(x,\;y)$ is a image-label pair, and $p$ are the predicted probabilities and $\mathbb{I}$ is the indicator function.

In the inference phase, we apply $\arg\max$ to the logits to get the sub-genre that is most probable for a slice of a track. Furthermore, if we need to predict the sub-genre of the entire track, the simplest method is to add a voting classifier on top of the predictions of slices. Other approaches include applying a weighted sum over the logits of slices of one track. The weights can even be set as trainable parameters, since some parts of a track like the intro can sometimes be less significant, while other parts like the drop would be crucial.

\begin{figure*}[!htbp]
    \centering 
      \includegraphics[width=.25\linewidth]{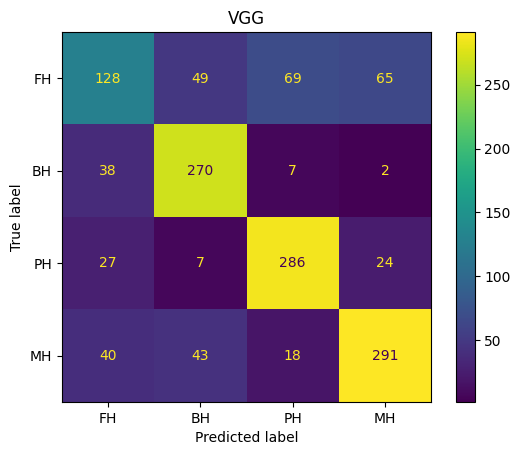}
      \label{fig:vgg_ori}
      \includegraphics[width=.25\linewidth]{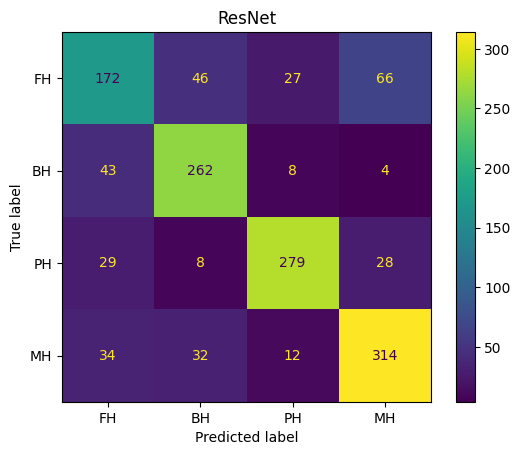}
      \label{fig:resnet_ori}
      \includegraphics[width=.25\linewidth]{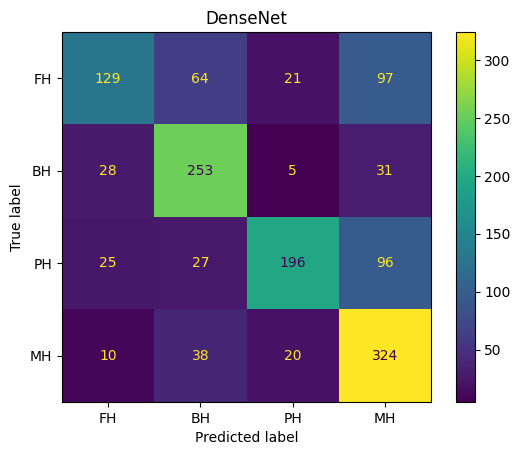}
      \label{fig:densenet_ori}
    \medskip
      \includegraphics[width=.25\linewidth]{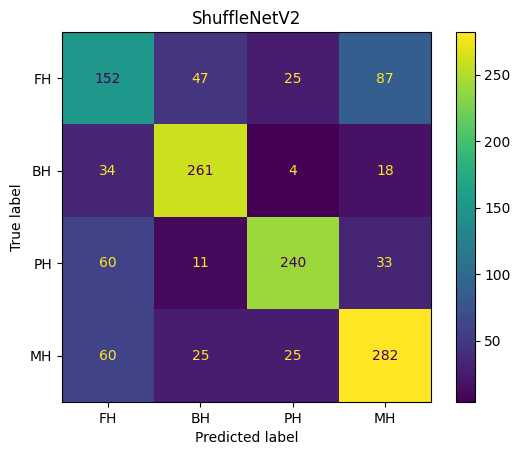}
      \label{fig:shufflenet_ori}
      \includegraphics[width=.25\linewidth]{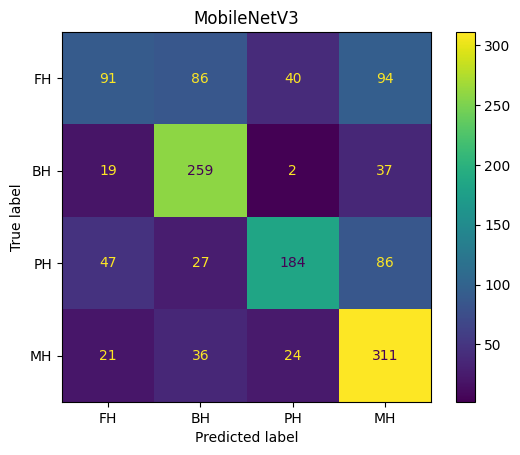}
      \label{fig:mobilenet_ori}
      \includegraphics[width=.25\linewidth]{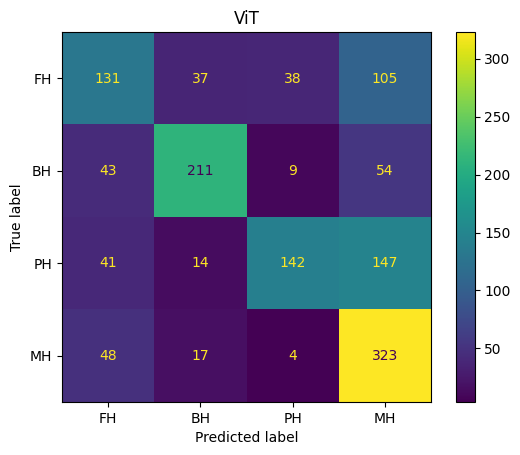}
      \label{fig:vit_ori}
    \caption{Confusion Matrices of the Test Set for Different Architectures}
    \label{fig:cm_ori} 
\end{figure*}

\begin{figure*}[!htp]
    \centering 
      \includegraphics[width=.25\linewidth]{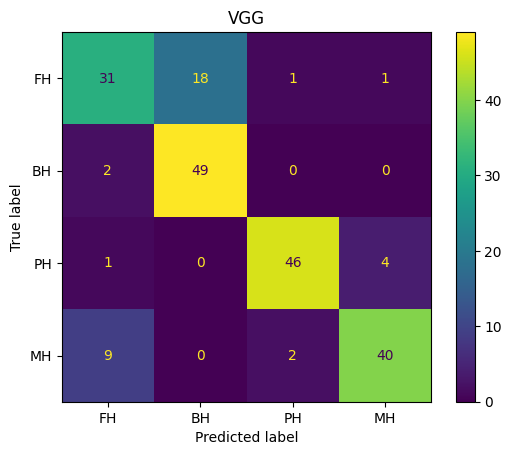}
      \label{fig:vgg_sub}
      \includegraphics[width=.25\linewidth]{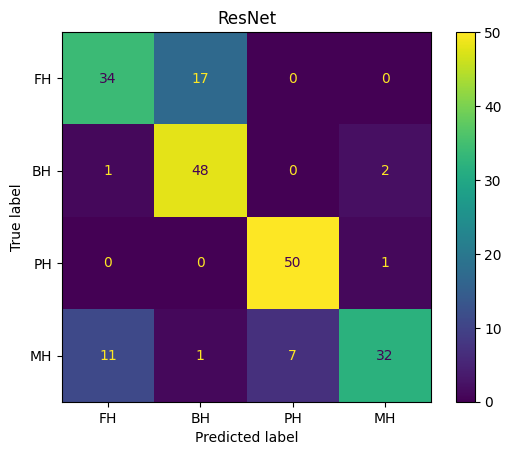}
      \label{fig:resnet_sub}
      \includegraphics[width=.25\linewidth]{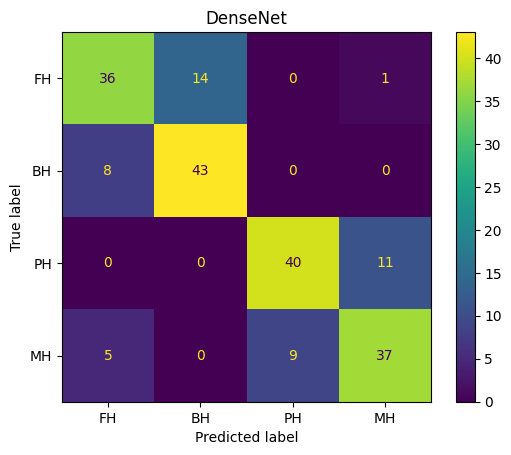}
      \label{fig:densenet_sub}
    \medskip
      \includegraphics[width=.25\linewidth]{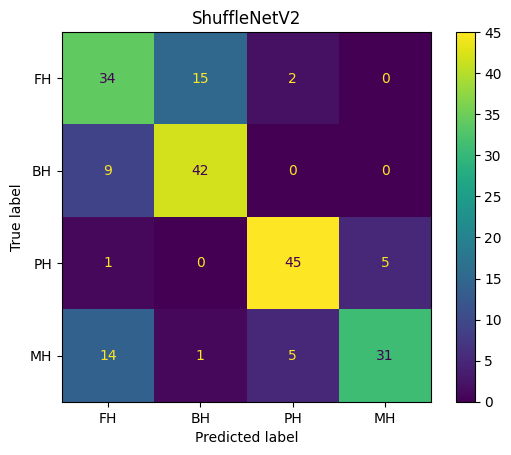}
      \label{fig:shufflenet_sub}
      \includegraphics[width=.25\linewidth]{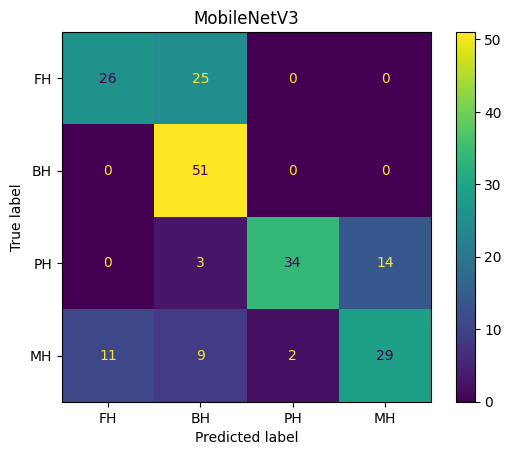}
      \label{fig:mobilenet_sub}
      \includegraphics[width=.25\linewidth]{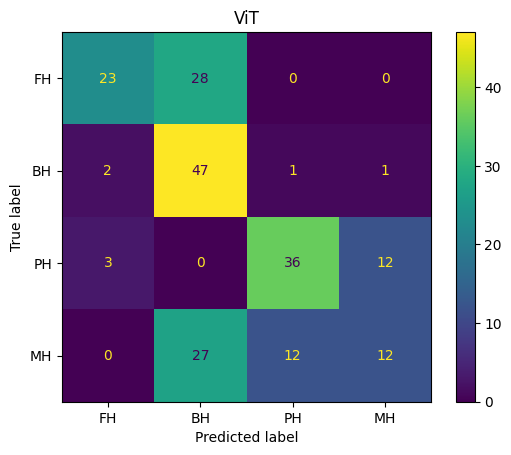}
      \label{fig:vit_sub}
    \caption{Confusion Matrices of the Test Set (drops only) for Different Architectures}
    \label{fig:cm_sub} 
\end{figure*}

\subsection{Neural Architecture}
We have tried several architectures, including 3 traditional convolutional neural networks \cite{vgg, resnet, densenet}, 2 lightweight convolutional neural networks \cite{shufflenet, mobilenet}  and a vanilla vision transformer \cite{vit} as backbones for the classification task. For these 5 CNN architectures, we put them before one dropout layer \cite{dropout} and one linear layer from $\mathbb{R}^{1000}$ to $\mathbb{R}^{1024}$. The input dimension is $1000$ since the ImageNet dataset \cite{imagenet} contains $1000$ categories. Besides, we have also train a $6$-layer vanilla vision transformer from scratch, which is built upon the attention mechanism instead of convolutional kernels. The details of our implementation can be viewed by the link provided in the introduction.

\section{Experiments}
We have conducted experiments using different types of neural networks as our backbone. As shown in Fig. \ref{fig:logs_ori} and \ref{fig:logs_sub}, all these neural architectures achieve competitive results for both the whole set and the drops-only subset.

From Fig. \ref{fig:cm_ori}, the first remark is that, it is challenging to distinguish future house tracks from melodic house tracks. Fortunately, even for human listeners, this dilemma can be tricky since some future house music possess a happy chord progression like those used in melody house tracks, while melody house tracks often have a synthesized lead which is bouncy and futuristic. Indeed, our classifiers have limited predictive power for these two sub-genres, but this is not necessarily bad. On the contrary, it is consistent with the fact that a single electronic track sometimes exhibit multiple styles.

Secondly, deeper and larger neural networks generally perform better than lightweight models. As listed in Tab. \ref{tab:results}, traditional deep architectures in our experiments, like ResNet \cite{resnet} and VGG \cite{vgg}, show better performance. This observation corresponds to our expectation that the timbre of sounds is a relatively high-level feature, which, on the spectrograms, should also need functions with stronger expressiveness to represent. For neural networks, deep architectures usually outperform wide ones, as shown by prior work \cite{deep_vs_wide_1, deep_vs_wide_2}.

Finally, comparing Fig. \ref{fig:cm_ori} and \ref{fig:cm_sub}, we could verify that the drops-only subset indeed shows significance with respect to the characteristics of each sub-genre. Fig. \ref{fig:cm_sub} shows that future house (FH) and bass house (BH) could be almost perfectly distinguished from progressive house (PH) and melodic house (MH), if we only focus on the drops of tracks. Such results are in line with the intrinsic features, or specifically, the use of sounds of these four sub-genres. A great common ground of future house and bass house is the massive use of synthesizer that makes a track sounds extremely electronic. On the contrary, acoustic (real) instruments are widely used in progressive house and melodic house tracks. For example, orchestral instruments like the violin are frequently used to create emotional atmosphere, while guitar strumming and piano chords often support the vocal, to give more emotions as well. Therefore, if we group future house and bass house as the "cool" category, and, group progressive house and melodic house as the "emotional" category, then models like VGG \cite{vgg} could achieve around $95\%$ accuracy on the classification of these two general categories. The classifiers perform better on the drops-only subset since this subset is "cleaner" in short. Beside the drop, other sections such as the intro of a track could exhibit variant peculiarities. For example, in the intro, future house tracks could include some bouncy piano chords, while melodic house tracks might contain some synthesized plucky sounds.

\section{Application Scenarios}
As mentioned in the introduction part, our work is specifically targeting at real-world application scenarios. In this section, we propose two cases where our dataset and models can be applied.
\subsection{Automation in Music Visualization}
A straightforward application is the automation of certain properties, such as the colors of lights, in the visualization of a mixtape. Normally, we want the real-time colors of the lights to reflect the atmosphere of the currently playing segment of the mixtape. Therefore, visual effects engineers would manually control the colors of lights throughout the whole mixtape. Using our model, the control of colors can be automated, which would bring more flexibility. For example, usually a DJ would fix the track list to be played before a show begins, to make sure the music matches up with the visual effects. Using our approach, the DJ performing onstage would be given more freedom to decide what to play next according to the reaction of the crowds, rather than the predetermined track list. We believe that such AI-assisted visualizer would bring much convenience for artists as a new paradigm.

As a result of our \href{https://www.wjx.cn/wxloj/datafullscreen.aspx?activity=173412412}{\color{blue}{survey}} on the correspondence between music genres and colors, we have pre-defined a mapping from sub-genres of house music, to triple tuples of colors, which could be modified in practice. Note that the following mapping is not completely the same as the one determined by choosing the top-$3$ colors for each sub-genre according to the survey. We have made slight modification on the colors of progressive house and future house to make the visualization more contrastive.

\begin{itemize}
    \item Future House: \textcolor{black}{blue}, \textcolor{black}{red}, \textcolor{black}{purple}.
\end{itemize}
\begin{itemize}
    \item Bass House: \textcolor{black}{black}, \textcolor{black}{white}, \textcolor{black}{purple}.
\end{itemize}
\begin{itemize}
    \item Progressive House: \textcolor{black}{orange}, \textcolor{black}{yellow}, \textcolor{black}{purple}.
\end{itemize}
\begin{itemize}
    \item Melodic House: \textcolor{black}{green}, \textcolor{black}{blue}, \textcolor{black}{yellow}.
\end{itemize}

\begin{figure}[!htbp]
    \centering
        \includegraphics[width=.22\textwidth]{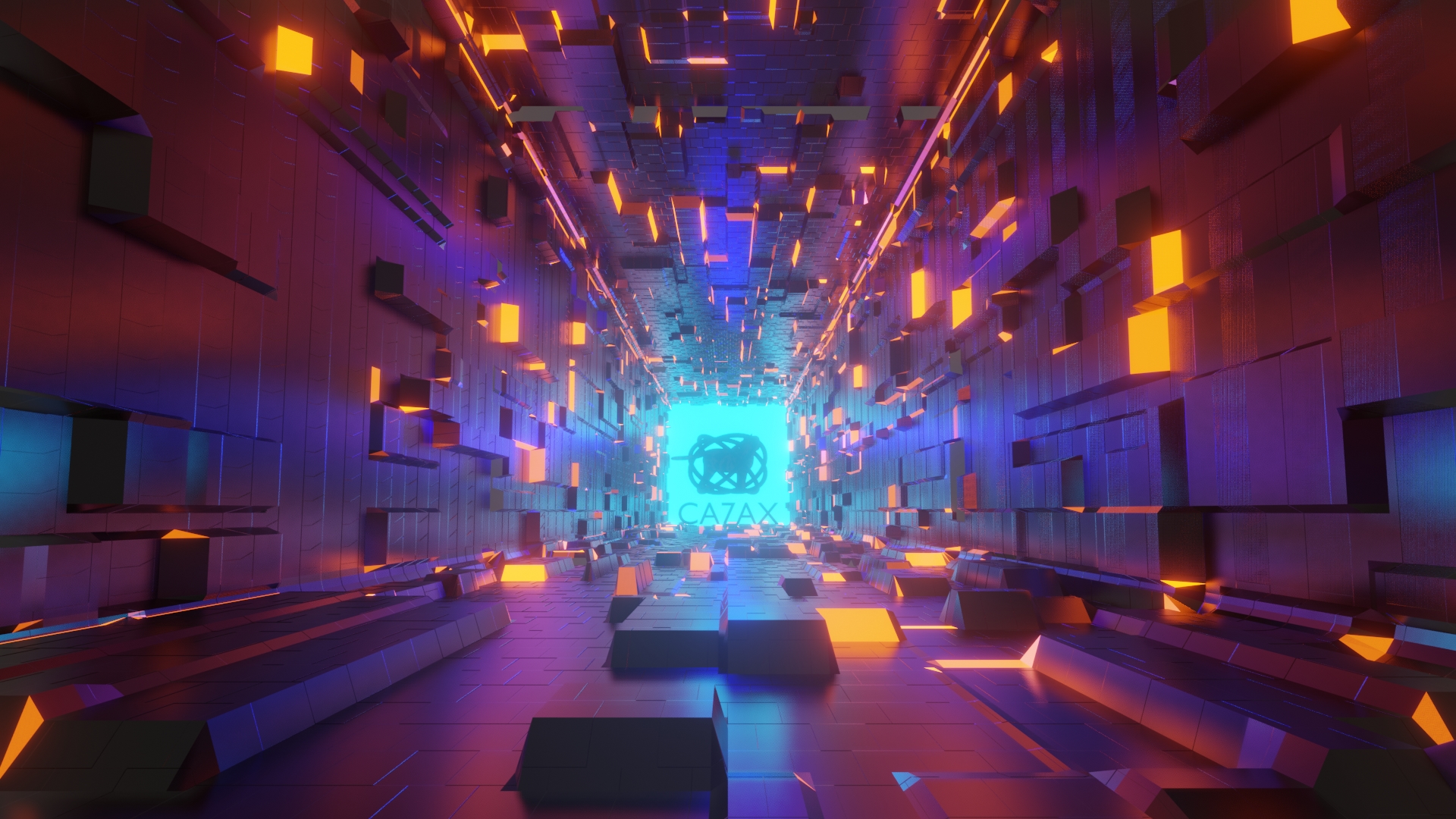}
        \includegraphics[width=.22\textwidth]{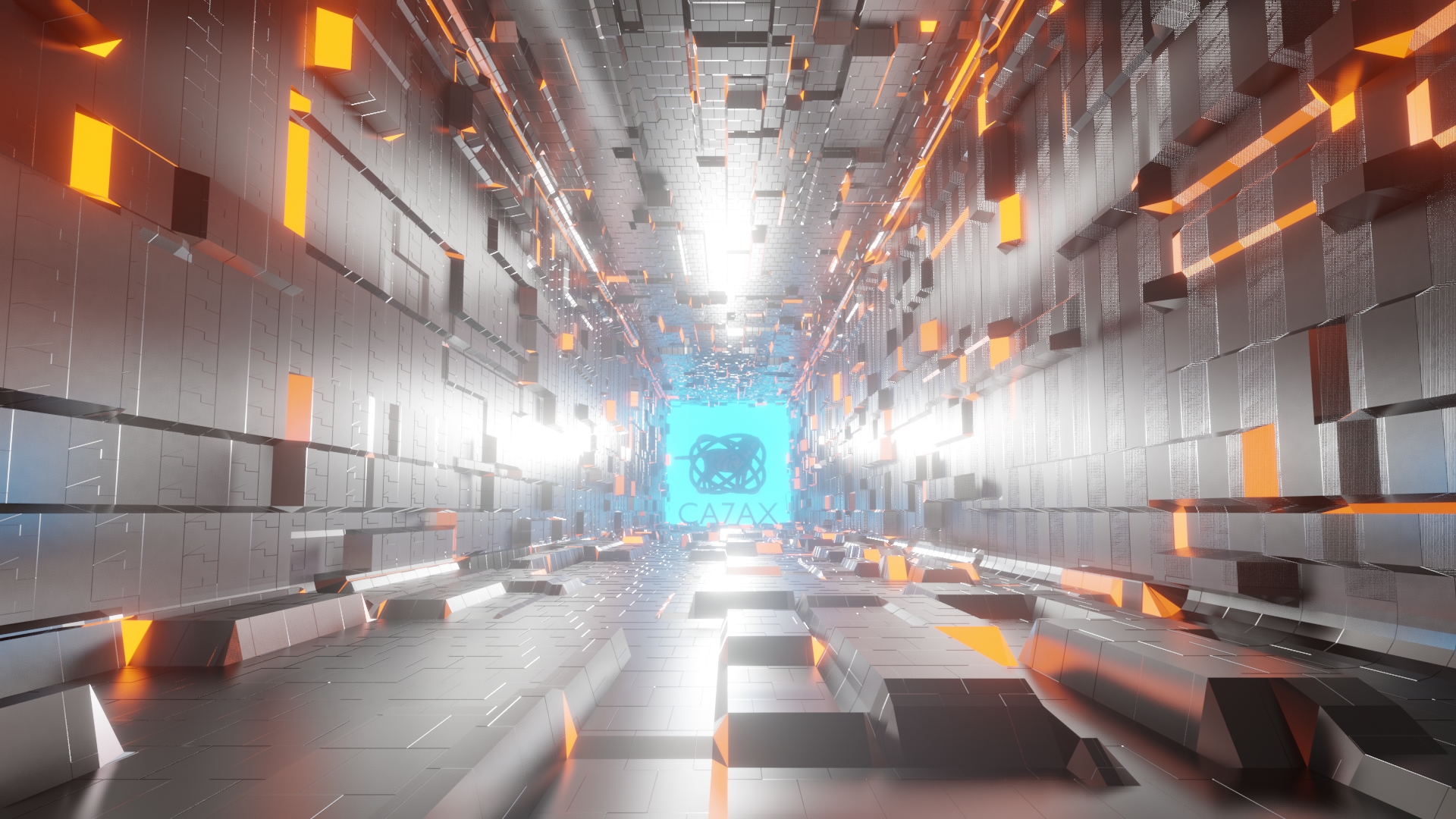}
    \par\smallskip
        \includegraphics[width=.22\textwidth]{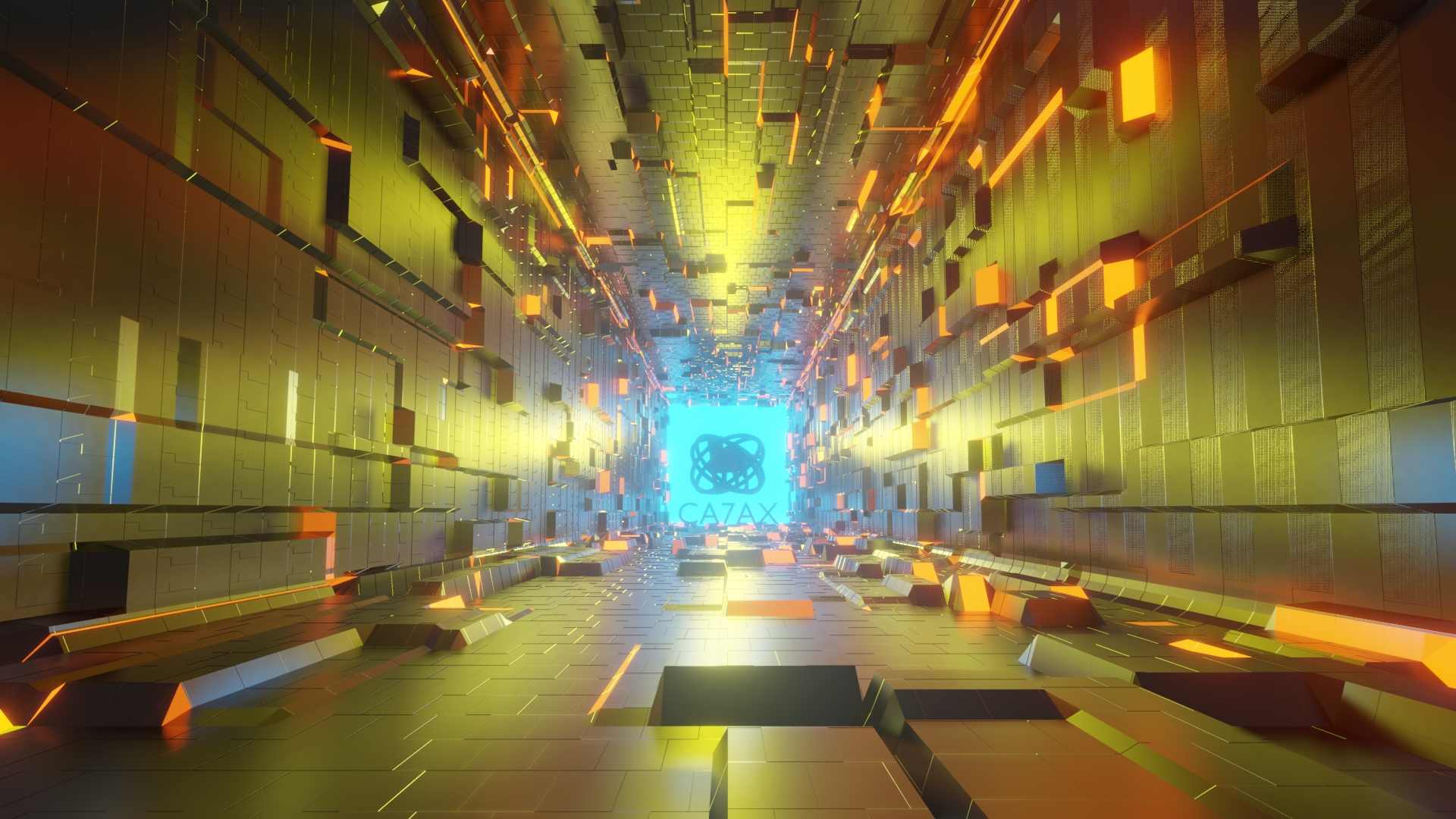}
        \includegraphics[width=.22\textwidth]{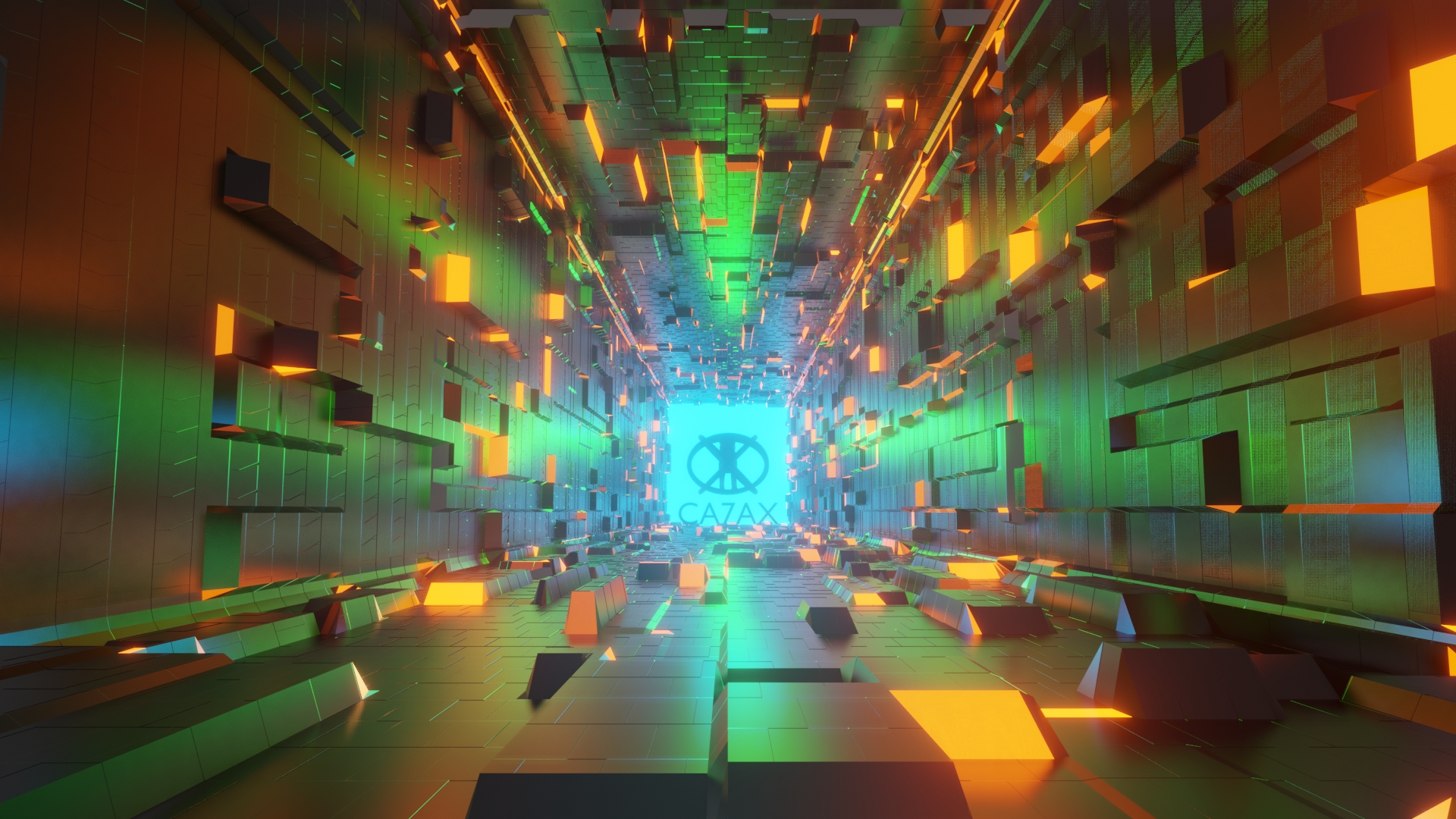}
\caption{Screenshots of Rendered Results}
\label{fig:screenshots}
\end{figure}

Then, after building a basic sci-fi environment in Blender \cite{blender}, we have rendered a demo for house music mixtape visualization, which can be viewed via this \href{https://drive.google.com/drive/u/1/folders/1Vmo9jMYbo0p1agjDoga7Onj1i4WbI4Wk}{\color{blue}{link}}. Note that this small set (mixtape) contains $4$ songs, mixed together in FL Studio 20 \cite{fl}. These $4$ songs are future house, bass house, progressive house and melodic house in chronological order. The screenshots of the rendered results are shown in Fig. \ref{fig:screenshots}.

\begin{figure}[!htbp]
    \centering
    \includegraphics[width=.41\textwidth]{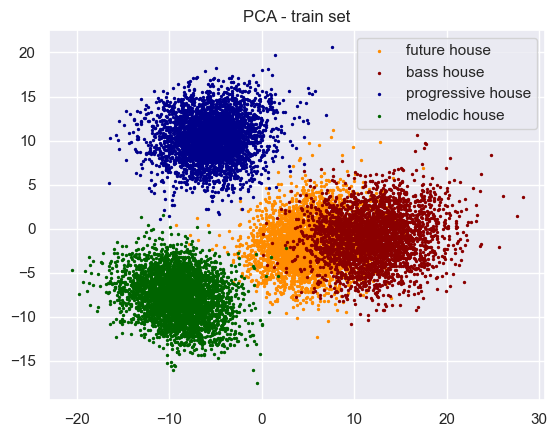}
    \includegraphics[width=.41\textwidth]{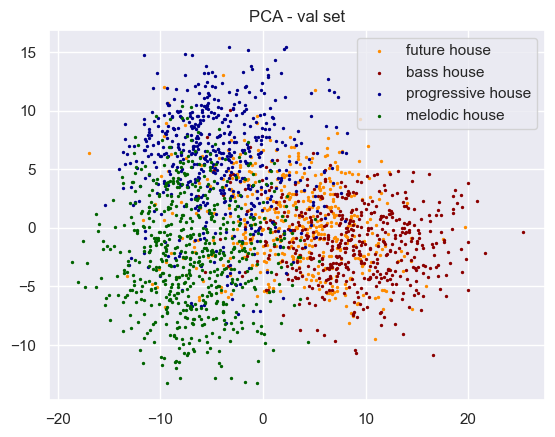}
    \includegraphics[width=.41\textwidth]{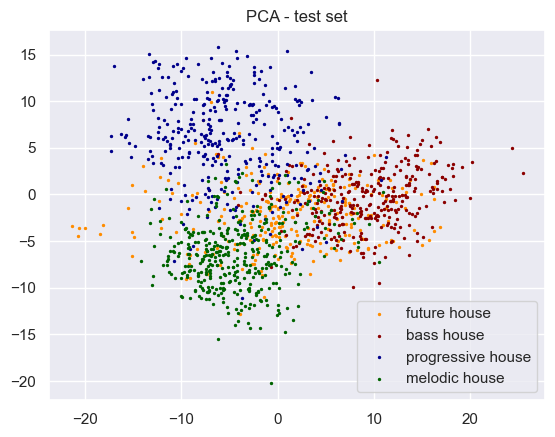}
\caption{2-PCA that Fits the Training Set}
\label{fig:pca}
\end{figure}

\subsection{Music Recommendation System}
Besides, this model can be integrated in fine-grained music recommendation system. Currently, many popular music streaming services have deployed recommendation systems for users to easily look for music that matches their appetite. With our approaches that extract information on the sub-genre level, the accuracy, or precisely, the level of customization of recommendation service could be brought to the next level.

\section{Future Works}
To sum up, we have annotated a house music dataset with sub-genre labels and built a baseline model for the classification task. Additionally, we have put forward a few real-world application scenarios where our work would be of great advantage. As shown in our demo videos, our approach could be considered as a new schema of music visualization. Still, there are several aspects that can be continued in the future.

Firstly, from the PCA analysis shown in Fig. \ref{fig:pca}, it seems that it is more challenging to distinguish future house from bass house, rather than melodic house. This does not perfectly match with our observation from the confusion matrices. Further investigation should be done to address this issue.

Secondly, there are other popular sub-genres of house music, like tropical house, to be included in our dataset. The reason why we have temporarily excluded it is that tropical house appear less frequently in music festivals compared to future house, progressive house, etc. Though less intense, tropical house tracks are so radio-friendly that they are also welcome by massive listeners. In view of such popularity, we might include dozens of tropical house tracks in our future work. 

Lastly, many other visual effects could also be automated by the information outputted by the model, such as the strength of the displacement of some geometry objects in a 3D environment. Hopefully, we would collaborate with 3D modeling experts to create visual effects that are more impactful and immersive.

\section*{Acknowledgment}

We thank members from New York University who have filled the questionnaire of the correspondence between colors and the music sub-genres, for our choices of colors in our demo. Besides, the Blender environment used in the demo is inspired by the work of Ducky 3D\cite{blender_base_env}. Also, we thank Wang et al. for assistance on the use of the high performance computing cluster (NYU HPC).

\bibliographystyle{ieeetr}
\bibliography{ref}

\end{document}